\documentclass[fleqn,usenatbib]{mnras}

\usepackage[labelfont={color=red}]{caption}
\usepackage{newtxtext,newtxmath}

\usepackage[T1]{fontenc}
\usepackage{tikz,xcolor,hyperref}

\definecolor{lime}{HTML}{A6CE39}
\DeclareRobustCommand{\orcidicon}{
\begin{tikzpicture}
\draw[lime, fill=lime] (0,0)
circle[radius=0.16]
node[white]{{\fontfamily{qag}\selectfont \tiny \.{I}D}}; 
\end{tikzpicture}
\hspace{-2mm}
}
\foreach \x in {A, ..., Z}{%
\expandafter\xdef\csname orcid\x\endcsname{\noexpand\href{https://orcid.org/\csname orcidauthor\x\endcsname}{\noexpand\orcidicon}}
}

\usepackage{orcidlink} 

\DeclareRobustCommand{\VAN}[3]{#2}
\let\VANthebibliography\thebibliography
\def\thebibliography{\DeclareRobustCommand{\VAN}[3]{##3}\VANthebibliography}

\usepackage{graphicx}	
\usepackage{amsmath}	
\usepackage{float} 
\usepackage{stfloats}
\usepackage{cite}
\usepackage{booktabs}
\usepackage{multirow}
\usepackage{xcolor}

\title[Properties of a Fading AGN from SDSS-IV MaNGA]{Properties of a Fading AGN from SDSS-IV MaNGA}

\author[Mo et al.]{Hao Mo\,\orcidlink{0000-0002-3443-0768}$^{1,2,3}$,
Yan-Mei Chen\,\orcidlink{0000-0003-3226-031X}$^{1,2,3}$\thanks{E-mail: chenym@nju.edu.cn},
Zhi-Yun Zhang\,\orcidlink{0009-0003-8089-3602}$^{1,2,3}$,
Alexei Moiseev\,\orcidlink{0000-0002-0507-9307}$^{4,5}$,
Dmitry Bizyaev\,\orcidlink{0000-0002-3601-133X}$^{5,6}$,\newauthor
Yong Shi\,\orcidlink{0000-0002-8614-6275}$^{1,2,3}$,
Qiu-Sheng Gu\,\orcidlink{0000-0002-3890-3729}$^{1,2,3}$,
Min Bao\,\orcidlink{0009-0005-9342-9125}$^{1,2,3}$,
Xiao Cao\,\orcidlink{0000-0002-8411-3314}$^{1,2,3}$
and Song-Lin Li\,\orcidlink{0000-0001-8112-7844}$^{7,8}$
\\
$^{1}$School of Astronomy and Space Science, Nanjing University, Nanjing 210093, China\\
$^{2}$Key Laboratory of Modern Astronomy and Astrophysics (Nanjing University), Ministry of Education, Nanjing 210093, China\\
$^{3}$Collaborative Innovation Center of Modern Astronomy and Space Exploration, Nanjing 210093, China\\
$^{4}$Special Astrophysical Observatory, Russian Academy of Sciences, Nizhny Arkhyz, 
Russia.\\
$^{5}$Sternberg Astronomical Institute, Lomonosov Moscow State University, Moscow, Russia. \\
$^{6}$Apache Point Observatory and New Mexico State University, Sunspot, NM, USA.\\
$^{7}$Research School of Astronomy and Astrophysics, Australian National University, Canberra, ACT 2611, Australia.\\
$^{8}$ARC Centre of Excellence for All-Sky Astrophysics in 3 Dimensions (ASTRO 3D), Australia.
}

\date{Accepted XXX. Received YYY; in original form ZZZ}

\pubyear{2022}

\begin{document}
\label{firstpage}
\pagerange{\pageref{firstpage}--\pageref{lastpage}}
\maketitle
\begin{abstract}
We identify a fading AGN SDSS J220141.64+115124.3 from the internal Product Launch-11 (MPL-11) in Mapping Nearby Galaxies at Apache Point Observatory (MaNGA) survey. The central region with a projected radius of $\sim$2.4 kpc is characterized as LINER-like line ratios while the outskirts extended to $\sim$15 kpc show Seyfert-like line ratios. The ${\hbox{[O\,{\sc iii}]}}$$\lambda$5007 luminosity of the Seyfert regions is a factor of 37 (2) higher than the LINER regions without (with) dust attenuation correction, suggesting that the AGN activity decreases at least $\sim$8 × 10$^3$ yrs ($\sim$2.4 kpc/light-speed) ago. We model the emission line spectra in the central region with double Gaussian components (a narrow core and a broad wing) and analyze the properties of each component. The narrow core component mostly co-rotates with the stellar disc, whereas the broad wing component with a median of the velocity dispersion $\sim$300 km s$^{-1}$ is related to a wind outflow. The kinematic position angle (PA) of the ionized gas shows a $\sim$20° twist from the galaxy center to 1.5 effective radius. The median of the PA difference between the gas and stellar components is as large as $\sim$50° within 0.4 effective radius. The tidal feature in DESI image and star-gas misalignment suggest this galaxy is a merger remnant. Combining all these observational results as well as public available X-ray and MIR luminosities, we confirm this is a fading AGN, the merger process kick-started the central engine to quasar phase which ionized gas composed of tidal debris, and now the activity of the central black hole decreases. The discontinuity in ${\hbox{[O\,{\sc iii}]}}$$\lambda$5007 flux and EQW maps is due to multiple AGN outbursts triggered by merger remnant gas inflows. 
\end{abstract}




\section{Introduction}
Active galactic nuclei (AGN) can have a strong influence on the interstellar medium of the host galaxies, through photoionization of the gas, the mechanical input from radio jets or winds driven by AGNs \citep{2017FrASS...4...42M}. These kinds of AGN feedback is thought to play a critical role in regulating the evolution of both the host galaxy and the central supermassive black hole (SMBH) \citep[e.g.,][]{2000ApJ...539L..13G, 2013ARA&A..51..511K}.\\
\indent The AGN can photoionize gas out to a scale of $\sim$1 kpc, which is the well-known narrow-line region (NLR). Observations also show that AGN-ionized gas can often extend well beyond this limit and reach tens of kiloparsecs \citep[e.g.,][]{2013MNRAS.436.2576L, 2014MNRAS.441.3306H, 2022ApJ...936...88F}. Extended emission line regions (EELRs) can exhibit complex morphologies and be spatially and kinematically distinct from the NLR. In early studies, the EELRs are connected with the presence of radio jets \citep{2006NewAR..50..694S}. However, following works find EELRs in active galaxies without radio emission \citep{2013A&A...549A..43H}. There are some EELRs show modest electron temperatures and narrow line widths which are much smaller than that generated from direct interaction with either an outflow or a radio jet, indicating they are ionized by radiation from the nucleus \citep[e.g.,][]{2020MNRAS.496.1035K}. Due to their physical extension, EELRs provide an ideal laboratory to probe the effect of the AGN on their galaxy \citep[e.g.,][]{2015ApJ...800...45H, 2017ApJ...835..222S}. Furthermore, we can obtain a view into the activity history of the central black holes through light-travel time to the gas clouds \citep{1994PASJ...46..539A, 2022ApJ...936...88F, 2023Galax..11..118M}.\\
\indent The suggestion that EELRs can be the result of a fading AGN was originally explored in an extended, highly ionized cloud called Hanny’s Voorwerp 15$\sim$35 kpc \citep{2012AJ....144...66K} to the south of galaxy IC 2497 \citep{2009MNRAS.399..129L}. It is bright in SDSS $g$-band due to unusually strong ${\hbox{[O\,{\sc iii}]}}$$\lambda$4959, 5007 emission lines. The difference in the energy budget between the nucleus and the ionization requirements in Hanny’s Voorwerp indicates that the AGN faded from quasar phase to a modest Seyfert or LINER level \citep{2017ApJ...835..256K}. Starting with the unusual morphology and ${g}$, ${r}$, ${i}$ colors of Hanny’s Voorwerp, similar objects were discovered from the Galaxy Zoo \citep{2012MNRAS.420..878K,2013PASP..125....2K} and from the SDSS survey \citep[“the green beans”,][]{2013ApJ...773..148S}. These are luminous ${\hbox{[O\,{\sc iii}]}}$ emitting clouds with extraordinarily high equivalent widths (EQWs) of several tens angstrom. More similar cases of extended ionized gas of lower luminosity AGNs have been reported using Integral Field Unit (IFU) spectroscopy in the local universe, e.g. an off-nuclear Seyfert-like compact emission line region in spiral galaxy NGC 3621 by \citet{2016ApJ...817..150M}, they suggest this emission line region as a “light echo” from an active galactic nucleus which has decreased by a factor of 13$\sim$500 during the last $\sim$230 yr; two asymmetric gas clouds have been found by \citet{2017ApJ...849..102C} in SDSS J1354+1327, one $\sim$10 kpc cone of photoionized gas to the south of the galaxy center and one $\sim$1 kpc semi-spherical front of shocked gas to the north of the galaxy center, they suggest the two clouds as a result of different AGN accretion cycles; \citet{2022ApJ...933..110X, 2023ApJ...943...28X} also found EELRs due to past AGN activities in NGC 7496 $\&$ NGC 5195. Mostly targeting normal galaxies, the MaNGA survey is ideal for searching low luminosity AGN echoes, which should be much more common than the extremely luminous AGN echoes identified from the SDSS imaging data, providing critical insights into the fueling, triggering, and starving of lower level AGN activity, as well as how the maintenance-mode AGN feedback operates in low-redshift galaxies. The first attempt of this sort of search was performed by \citet{2023ApJ...950..153F} who found several EELRs related to fading AGNs in MaNGA sample of post-starburst galaxies.\\
\indent SDSS J220141.64+115124.3 (hereafter SDSS J2201+1151), the galaxy studied in this work, was first considered as a fading AGN candidate in \citet{2012MNRAS.420..878K} along with other 18 galaxies with AGN ionized regions at projected radii > 10 kpc. They estimate a ratio of 3.4 between the ionizing luminosity of the EELRs and far-infrared (FIR) AGN luminosity, indicating the extended cloud can not be ionized by an obscured AGN in SDSS J2201+1151. The most direct interpretation is that the EELRs are ionized by AGN that has faded over the light-travel time between the ionized gas and nucleus. These results were confirmed in the study of ionization history based on HST narrow band imaging \citep{2017ApJ...835..256K}. The ionized gas kinematics of this galaxy was briefly considered by fitting ${\hbox{[O\,{\sc iii}]}}$$\lambda$5007 emission line using a single Voigt-profile \citep[the convolution of the Gaussian and Lorentz profiles,][]{2008AstBu..63..181M} based on data from Big Telescope Alt-azimuthal (BTA) in its scanning Fabry-Perot Interferometer mode \citep[FPI,][]{2015AJ....149..155K}.
\\
\indent In this work, we use the MaNGA data to study the gas and stellar components in SDSS J2201+1151. The paper is organized as follows. In Section 2, we give a short introduction to the MaNGA survey, general properties of this galaxy, and method of emission line fitting. In Section 3, we analyze the detailed properties of this galaxy, including the velocity field, ${\hbox{[S\,{\sc ii}]}}$-BPT diagram, dust attenuation, AGN luminosity, and kinematic position angle of the gas and stellar components. We discuss the observational results in Section 4. Finally, we draw conclusions in Section 5. Throughout this paper, we use the flat $\Lambda$CDM cosmological parameters with $H_{0}$ = 70km s$^{-1}$ Mpc$^{-1}$, $\Omega_{m} = 0.3$, and $\Omega_{\Lambda} = 0.7$. All wavelengths stated are in vacuum.

\section{Observations $\&$ Data Reduction}
\subsection{The MaNGA survey}
MaNGA is one of three core programs in the fourth-generation Sloan Digital Sky Survey \citep[SDSS-IV,][]{2017AJ....154...28B,2016AJ....151....8Y} started on July 2014, using 2.5m Sloan Foundation Telescope at Apache Point Observatory \citep[]{2006AJ....131.2332G}. MaNGA employs dithered observations with 17 fiber-bundle IFUs \citep{2015AJ....150...19L} with 5 sizes varying between 19 fibers and 127 fibers (or 12.5${''}$-32.5${''}$ diameter in the sky) to explore the detailed structure of nearby galaxies \citep{2015AJ....149...77D}. Two dual-channel BOSS spectrographs \citep{2013AJ....146...32S} provide simultaneous spectral coverage over 3,600$\sim$10,300{\AA} at a median resolution of $R$ $\sim$2,000. MaNGA selects "Primary" and "Secondary" samples defined by two radial coverage goals. The "Primary" sample is extended out to $\sim$1.5 effective radius ($R_{e}$, Petrosian 50$\%$ light radius) while the "Secondary" sample is extended out to $\sim$2.5 $R_{e}$ \citep{2016AJ....152..197Y}. SDSS J2201+1151 belongs to the "Primary" sample and has $R_{e}$ of $\sim$10${''}$ in $r$-band. The 2${''}$ fiber diameter corresponds to 1.2kpc spatial resolution at the redshift $z$ $\sim$0.0298 of this galaxy. A typical exposure time of 3 hours on-sky ensures a per-fiber $r$-band continuum signal-to-noise ratio (S$/$N) of 5 at 1.5$R_{e}$, with much higher S$/$N $\sim$100 towards the center. 

The Data Reduction Pipeline \citep[DRP,][]{2016AJ....152...83L} of MaNGA provides sky-subtracted and flux-calibrated 3D spectra for each galaxy. The Data Analysis Pipeline \citep[DAP,][]{2019AJ....158..231W} is a survey-led software package for analyzing the spectra produced by the DRP.
It has been developed since 2014 and has gone through several versions, which heavily relies on pPXF \citep{2004PASP..116..138C} and a subset of stellar templates drawn from the MaSTar library \citep{2019ApJ...883..175Y}. DAP fits the stellar continuum in each spaxel and produces estimates of the lick indexes and measurements of 21 major nebular emission lines in the MaNGA wavelength coverage. In this work, we use the following parameters from the MaNGA DAP products drawn from the internal MaNGA Product Launch-11 (MPL-11) named "MAPS-SPX-MILESHC-MASTARSSP", including line-of-sight rotation velocity of stars ($ V\rm_{\star}$), stellar velocity dispersion ($\rm \sigma_{\star}$), line-of-sight rotation velocity of ionized gas ($ V\rm_{gas}$), gas velocity dispersion ($\rm \sigma_{gas}$), ${\hbox{[O\,{\sc iii}]}}$$\lambda$5007 flux, ${\hbox{[O\,{\sc iii}]}}$$\lambda$5007 equivalent width (EQW).

\subsection{General properties of SDSS J2201+1151}
\begin{figure*}
	\centering
	\includegraphics[width=\textwidth]{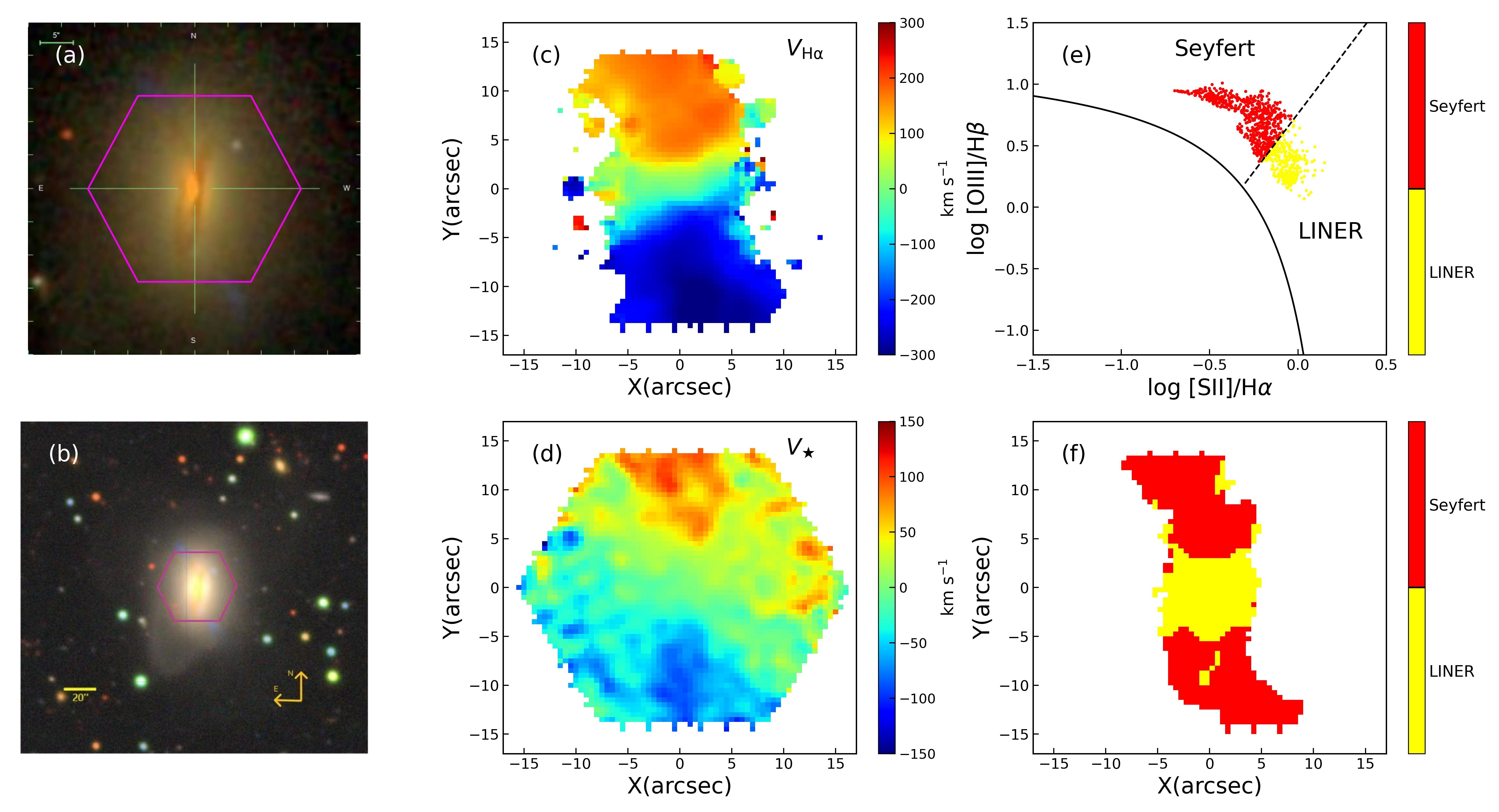}
	\caption{(a) The sloan ${g}$, ${r}$, ${i}$ color image of SDSS J2201+1151. (b) The DESI image which is $\sim$2 mag deeper than SDSS. In both panel (a) and (b), the purple hexagon marks the region covered by the MaNGA bundle. (c) The ionized gas velocity field traced by H$\alpha$ emission line, only spaxels with H$\alpha$ S$/$N larger than 3 are included.  
    (d) The velocity field of the stellar component. Red represents moving away from us while blue represents moving toward us, and the color bars indicate the value of velocities.
    (e) The ${\hbox{[S\,{\sc ii}]}}$-BPT diagram. The black solid curve marks the theoretical upper boundary for extreme starbursts determined by \citet{2001ApJ...556..121K}. The black dashed line is the division between LINER and Seyfert regions \citep{2006MNRAS.372..961K}. Yellow represents the LINER region and red represents the Seyfert region.
    (f) The spatially resolved BPT diagram. The color definition is the same as panel (e).} 
	\label{1}
\end{figure*}
\indent Figure \ref{1} shows the gas \& stellar kinematics, as well as ionization mechanism of SDSS J2201+1151. Figure \ref{1}(a) is the SDSS ${g}$, ${r}$, ${i}$ color image in which the purple hexagon marks the region covered by the MaNGA bundle. Figure \ref{1}(b) is the DESI image which is $\sim$2 mag deeper than SDSS, the tidal feature in the bottom left of this galaxy was first discovered in the Big Telescope Alt-azimuthal (BTA) $R$-band image presented by \citet{2015AJ....149..155K}, again the purple hexagon marks the region covered by the MaNGA bundle. Figure \ref{1}(c) shows the ionized gas velocity field traced by the H${\alpha}$ emission line, only spaxels with H${\alpha}$ S$/$N larger than 3 are included. 

Figure \ref{1}(d) shows the velocity field of the stellar component. Red represents moving away from us while blue represents moving toward us, and the color bars indicate the value of velocities. Figure \ref{1}(e)$\&$(f) show the ${\hbox{[S\,{\sc ii}]}}$-BPT diagnostic diagram \citep{1981PASP...93....5B,1987ApJS...63..295V} as well as the spatial resolved BPT for spaxels with S$/$N larger than 3 for the four emission line (H${\beta}$, ${\hbox{[O\,{\sc iii}]}}$$\lambda$5007, H${\alpha}$, ${\hbox{[S\,{\sc ii}]}}$$\lambda$$\lambda$6717,6731 doublets). The solid curve in Figure \ref{1}(e) marks the theoretical upper boundary of extreme starbursts determined by \citet{2001ApJ...556..121K}. The black dashed line is the division between low-ionization nuclear emission-line regions (LINERs, below) and Seyfert (above) galaxies \citep{2006MNRAS.372..961K}. In Figure \ref{1}(e)$\&$(f), yellow represents the LINER region and red represents the Seyfert region. It is totally out of our expectation that the central region of this object is characterized as LINER-like line ratios while the outskirts show Seyfert-like line ratios.\\
\begin{figure*}
	\centering
		\includegraphics[width=0.8\textwidth]{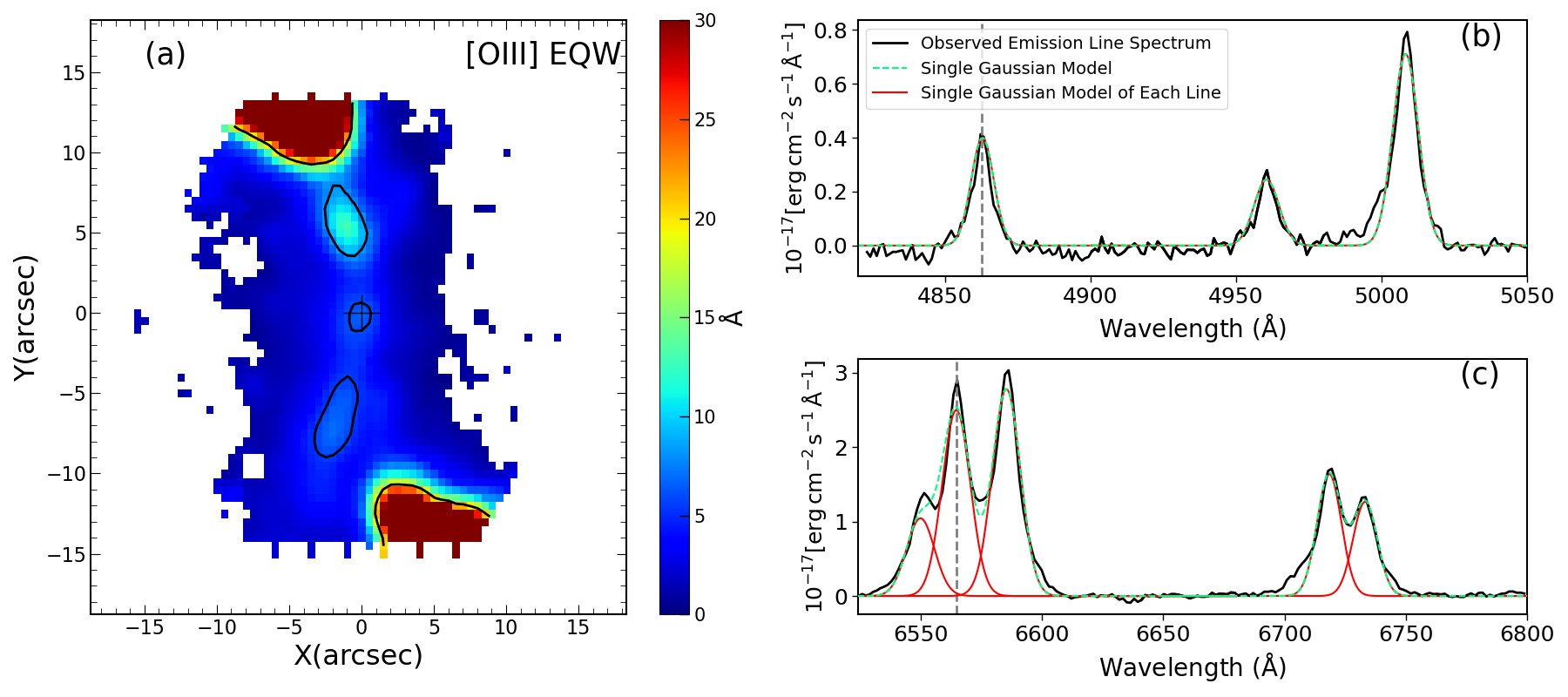}
	\caption{(a) The ${\hbox{[O\,{\sc iii}]}}$$\lambda$5007 equivalent width map. The galaxy center is marked with a black cross while the black contours mark the regions with enhanced ${\hbox{[O\,{\sc iii}]}}$$\lambda$5007 EQW. (b)$\&$(c) The central emission line spectrum of the H${\beta}$ $\&$ H${\alpha}$ regions, respectively. Black is the observed emission line spectrum, red is the single Gaussian model of each emission line given by MaNGA DAP while green is the combination of all the Gaussian components. The vertical gray lines mark the restframe wavelength center of the H${\beta}$ and H${\alpha}$ emission lines.}
	\label{2}
\end{figure*}

\begin{figure*}
	\centering
		\includegraphics[width=0.8\textwidth]{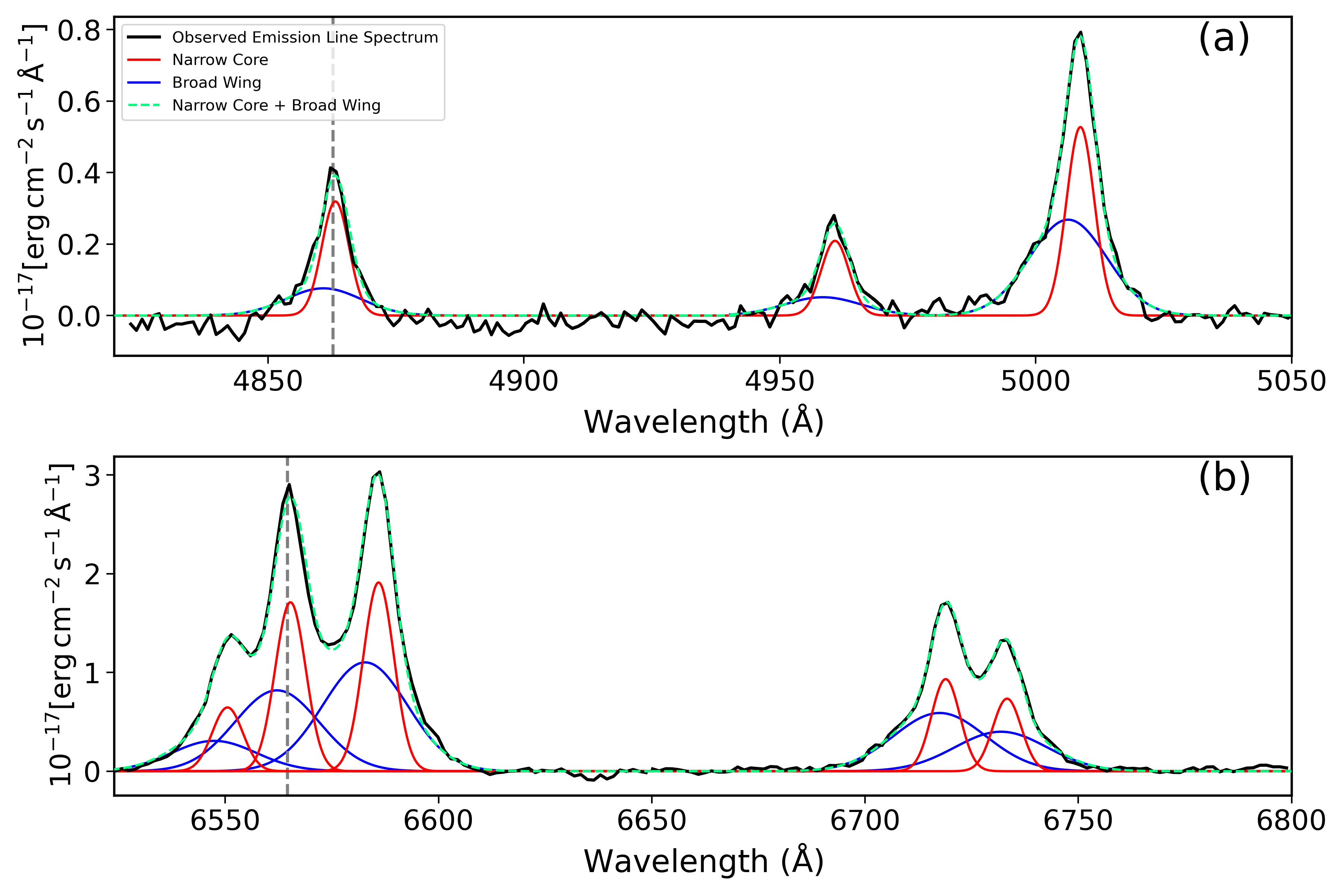}
	\caption{The central emission line spectrum of the H${\beta}$ (top) and H${\alpha}$ (bottom) regions.
    Black is the observed emission line spectrum,
    red represents the narrow core component while blue represents the broad wing component. The green spectrum is the combination of the blue and red components, which is the best-fit model. The vertical gray lines mark the rest-frame wavelength center of H${\beta}$ and H${\alpha}$ emission lines.}
	\label{3}
\end{figure*}

\subsection{Method of emission line fitting} \label{2.3}
\indent In order to explore the reliability of ionization state measurements and the origin of the ionization state variations from the center to the outskirts, we check the spectrum of each spaxel, finding that in some spaxels the emission line spectra can not be modeled by a single Gaussian component, especially for the central region. Figure \ref{2} shows the emission line spectrum for the central spaxel. Figure \ref{2}(a) shows the ${\hbox{[O\,{\sc iii}]}}$$\lambda$5007 EQW map. The galaxy center is marked with a black cross while the black contours mark the regions with enhanced ${\hbox{[O\,{\sc iii}]}}$$\lambda$5007 EQW. Figure \ref{2}(b)$\&$(c) show the center emission line spectrum in the H${\beta}$ and H${\alpha}$ regions, respectively. Black is the observed emission line spectrum, red is the single Gaussian model given by MaNGA DAP while green is the combination of all the Gaussian components. It is obvious that a single Gauss is insufficient for describing the profile of the emission lines.
\\ 
\indent In this section, we fit each emission line (H${\beta}$, ${\hbox{[O\,{\sc iii}]}}$$\lambda$4959, 
${\hbox{[O\,{\sc iii}]}}$$\lambda$5007, ${\hbox{[N\,{\sc ii}]}}$$\lambda$6548, H${\alpha}$, ${\hbox{[N\,{\sc ii}]}}$$\lambda$6583, ${\hbox{[S\,{\sc ii}]}}$$\lambda$$\lambda$6717, 6731) with double Gaussian components (a narrow core plus a broad wing) using the package curve\_fit in SciPy \citep{2020SciPy-NMeth}. Since each Gaussian component of these lines likely arises from the same physical region with similar kinematics, H${\beta}$ and ${\hbox{[O\,{\sc iii}]}}$$\lambda$4959 are tied to have the same line center and line width as ${\hbox{[O\,{\sc iii}]}}$$\lambda$5007 in the velocity space for each Gaussian component, while ${\hbox{[N\,{\sc ii}]}}$$\lambda$$\lambda$6548,6583 and ${\hbox{[S\,{\sc ii}]}}$$\lambda$$\lambda$6717,6731 are tied to have the same line center and line width as H${\alpha}$. We also tried to untie the line center and line width of the emission line components, finding our following results keep the same.
\\
\indent In Figure \ref{3}, we show the emission line spectrum of the central spaxel as well as its best-fitting double Gaussian model as an example. Black is the observed emission line spectrum, red represents the narrow core components while blue represents the broad wing components. Green is the combination of the blue and red components, which is the best-fit model. Comparing with the single Gaussian model given by MaNGA DAP in Figure \ref{2}, it is obvious that the double Gaussian describes the spectrum much better than the single Gaussian model.\\
\indent In order to qualify which spaxels require double Gaussian models, we calculate the reduced chi-square ($\chi^{2}$) of the single ($\chi^{2}_{\rm{1}}$) and double Gaussian models ($\chi^{2}_{\rm{2}}$) respectively. The reduced $\chi^{2}$ is defined as $\chi^{2}$ per degree of freedom \citep{1993stp..book.....L}. We apply the ratio of reduced ${\chi^2}$ between single and double Gaussian models ($\chi^{2}_{\rm{1}}$$/$$\chi^{2}_{\rm{2}}$) to qualify how significant the fitting is improved by the double Gaussian model \citep{2014RAA....14..913P}.\\ 
\begin{figure*}
	\centering
		\includegraphics[width=0.8\textwidth]{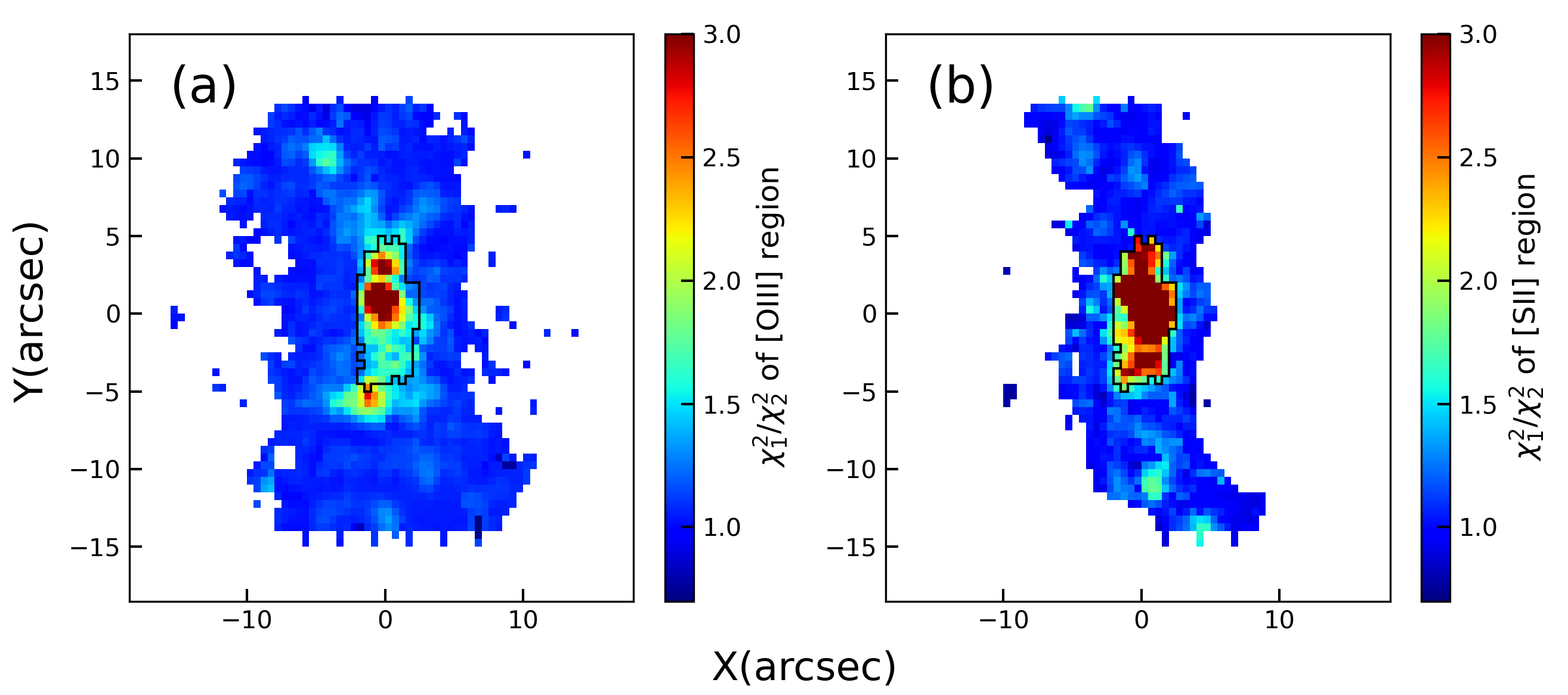}
	\caption{The ratio of $\chi^2$ between single and double Gaussian models.
    (a)$\&$(b) The reduced $\chi^2$ ratio of the ${\hbox{[O\,{\sc iii}]}}$$\lambda$5007 and 
    {\hbox{[S\,{\sc ii}]}}$\lambda$$\lambda$6717,6731 emission line regions, respectively.
    Spaxels within the black polygon have reduced $\rm{\chi^2}$ ratio larger than 1.2 for both ${\hbox{[O\,{\sc iii}]}}$$\lambda$5007 and {\hbox{[S\,{\sc ii}]}}$\lambda$$\lambda$6717,6731 emission lines.}
	\label{4}
\end{figure*}
\indent Figure \ref{4}(a) shows the reduced $\chi^2$ ratio of ${\hbox{[O\,{\sc iii}]}}$$\lambda$5007 region (4980\AA \, $\leq$  $\lambda$ $\leq$ 5030\AA) for spaxels with {\hbox{[O\,{\sc iii}]}}$\lambda$5007 S$/$N larger than 3. Figure \ref{4}(b) shows the reduced $\chi^2$ ratio of {\hbox{[S\,{\sc ii}]}}$\lambda$$\lambda$6717,6731 region (6690\AA \, $\leq$ $\lambda$ $\leq$ 6760\AA) for spaxels with {\hbox{[S\,{\sc ii}]}}$\lambda$$\lambda$6717,6731 S$/$N larger than 3. Spaxels with $\chi^{2}_{\rm{1}}$$/$$\chi^{2}_{\rm{2}}$ greater than 1.2 for both the ${\hbox{[O\,{\sc iii}]}}$$\lambda$5007 and {\hbox{[S\,{\sc ii}]}}$\lambda$$\lambda$6717,6731 emission lines (the region within the black polygon) are selected for the double Gaussian fitting \citep{2008ApJ...683L.115H}. We also visually inspect the emission line spectra by eyes to select spaxels requiring double Gaussian fitting for the emission lines. 90$\%$ spaxels selected by eyes are also selected by the ratio of reduced ${\chi^2}$ method. Overall, we use the double Gaussian model (a narrow core plus a broad wing) to describe the profile of the emission line spectra for spaxels within the black polygon. For the outer regions, a single Gaussian model (narrow core) is applied to fit each emission line.

\section{RESULTS}\label{result}

\subsection{The kinematics of gas $\&$ stellar components}
\begin{figure*}
	\centering
		\includegraphics[width=\textwidth]{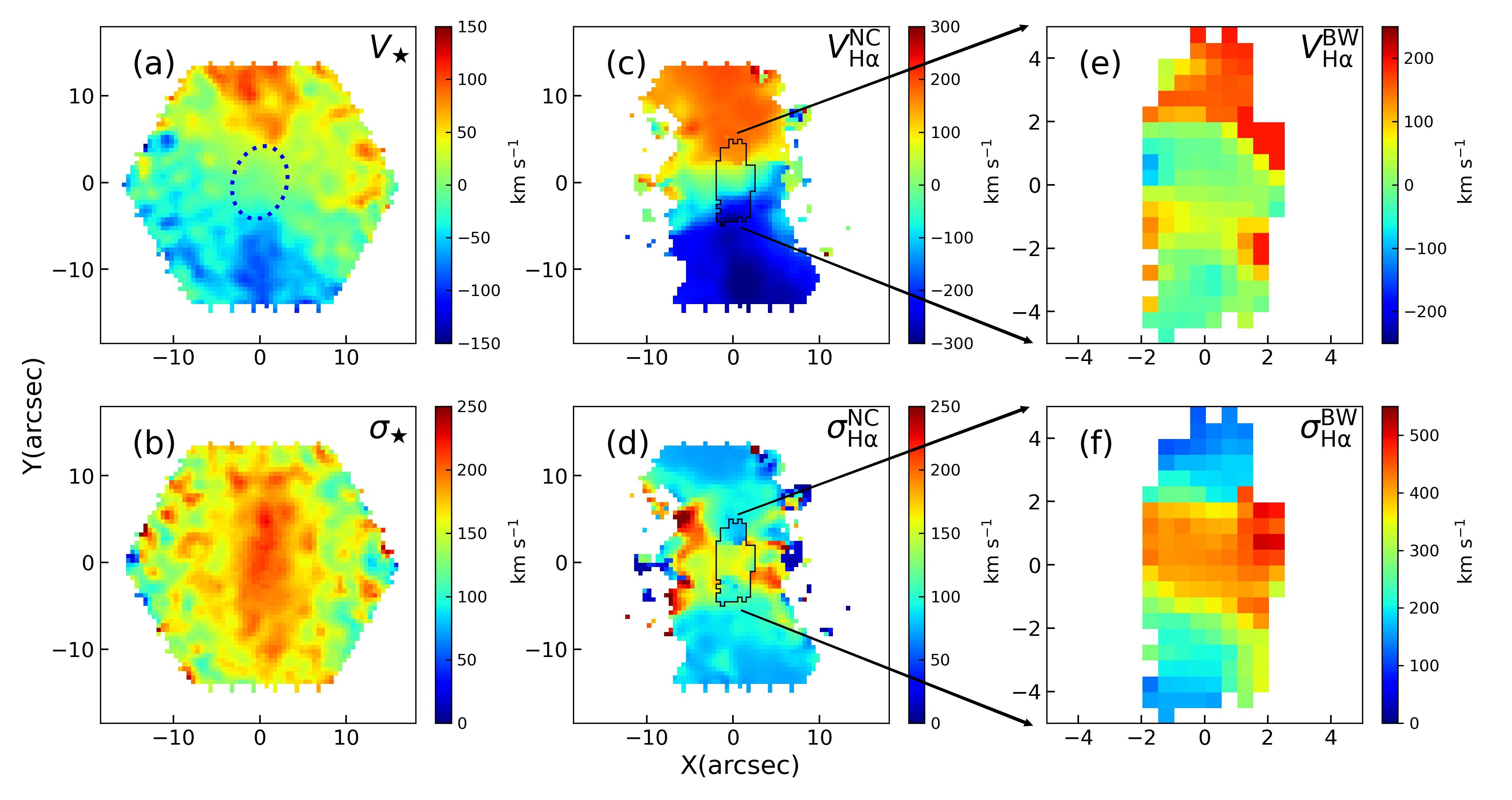}
	\caption{Kinematics for stellar and gas components.
    (a)$\&$(b) The stellar velocity and velocity dispersion fields given by MaNGA DAP. (c)$\sim$(f) The gas velocity and velocity dispersion fields of the narrow core (c,d) and broad wing (e,f) components.
    The blue dotted ellipse in panel (a) marks 0.4$R_{e}$ region.
    Similar to Figure \ref{4}, the black polygon marks the region which requires double Gaussian model.
    }
	\label{5}
\end{figure*}

In this section, we study the kinematics of stellar and gas components. The stellar velocity and velocity dispersion fields are given by MaNGA DAP. The gas velocity is estimated through comparing the restframe line center of an emission line to the line center of the best-fit Gaussian model and the gas velocity dispersion is calculated from the best-fit line width. Figure \ref{5}(a) shows the stellar velocity field. Figure \ref{5}(c)$\&$(e) show the gas velocity fields of the narrow core ($V$$^{\rm{NC}}_{\mathrm{gas}}$) and broad wing components ($V$$^{\rm{BW}}_{\mathrm{gas}}$), respectively. The narrow core component appears to co-rotate with the stellar component, while the broad wing component seems different from the stellar component and much more complicated. Figure \ref{5}(b) shows the stellar velocity dispersion, while Figure \ref{5}(d)$\&$(f) show the gas velocity dispersion of the narrow core ($\sigma$$^{\rm{NC}}_{\mathrm{gas}}$) and broad wing ($\sigma$$^{\rm{BW}}_{\mathrm{gas}}$) components, respectively. The median of the velocity dispersion is $\sim$100 km s$^{-1}$ for the narrow core component and $\sim$300 km s$^{-1}$ for the broad wing component. The velocity dispersion of the broad wing component within the central $\pm$2${''}$ region is up to 400$\sim$500 km s$^{-1}$, which is possibly related to a wind outflow. We discuss this possibility in more detail in Section \ref{dis1}. We note the fitting of H${\beta}$ regions gives similar velocity and velocity dispersion of narrow core $\&$ broad wing components to that of H${\alpha}$ region although they are not tied together in our fitting process.

\subsection{The ${\hbox{[S\,{\sc ii}]}}$-BPT diagram}
\begin{figure*}
	\centering
		\includegraphics[width=0.75\textwidth]{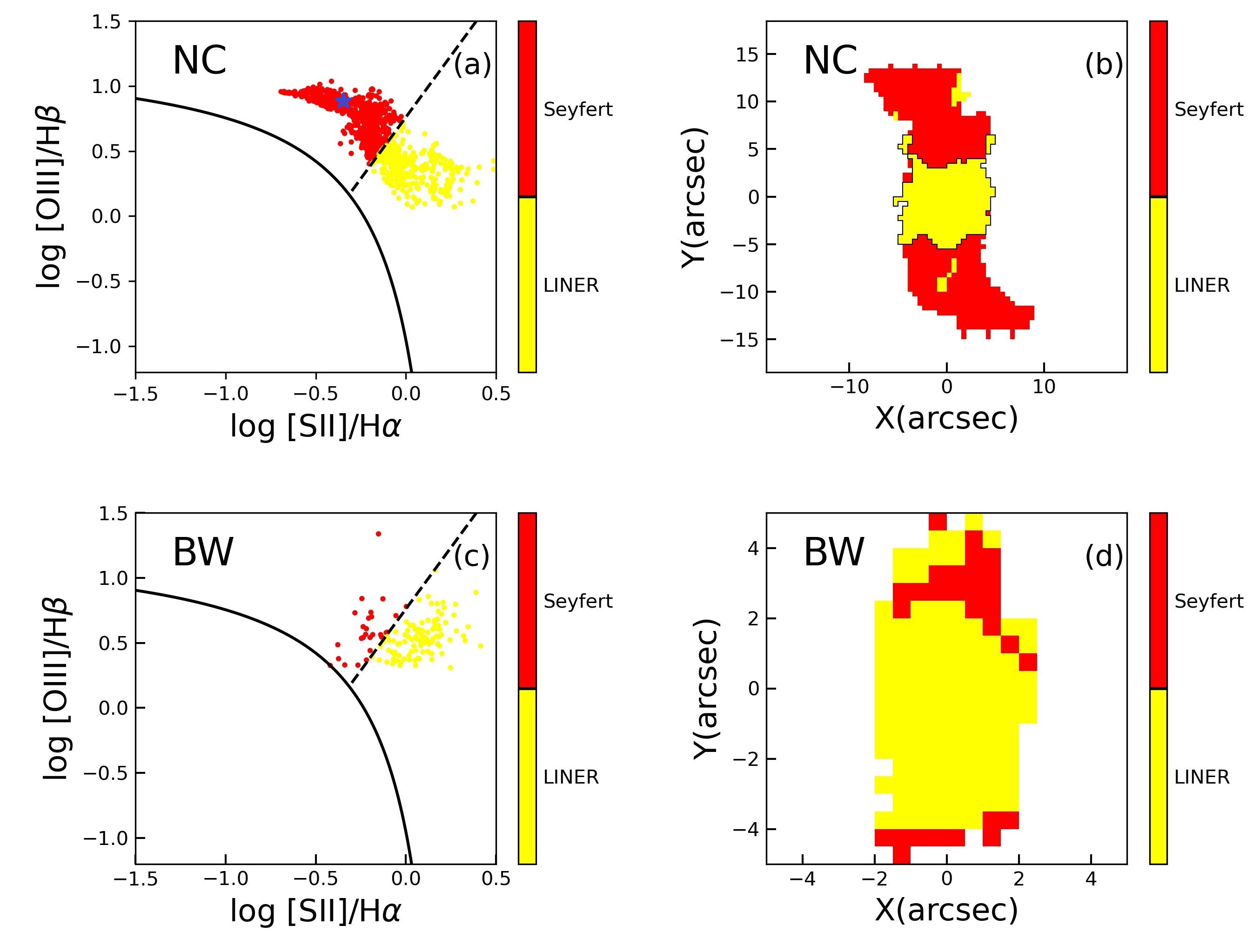}
	\caption{${\hbox{[S\,{\sc ii}]}}$$/$H${\alpha}$ vs. ${\hbox{[O\,{\sc iii}]}}$$/$H${\beta}$ diagnostic diagrams. 
    (a)$\&$(c) The ${\hbox{[S\,{\sc ii}]}}$-BPT diagrams of the narrow core and broad wing components, respectively. The blue star in panel (a) is the median line ratio measured from APO long-slit outside the MaNGA bundle.
    (b)$\&$(d) The spatially resolved BPT diagram of the narrow core and broad wing components. 
    The region within the black polygon is defined as LINER regions.}
	\label{6}
\end{figure*}

Mapping diagnostic line ratios across a galaxy reveals spatial variations in the ionization state of the gas \citep{2015ApJS..221...28R, 2016MNRAS.462.1616D, 2021ApJ...916...50F}. In this section, we re-visit BPT diagram using the Gaussian fitting results given in Section \ref{2.3}. Standard BPT diagnostic diagrams are based on the line ratios of ${\hbox{[O\,{\sc iii}]}}$$\lambda5007$$/$H${\beta}$ vs. ${\hbox{[N\,{\sc ii}]}}$$\lambda6583$$/$H${\alpha}$, ${\hbox{[S\,{\sc ii}]}}$$\lambda\lambda6717, 6731$$/$H${\alpha}$ or ${\hbox{[O\,{\sc i}]}}$$\lambda6300$$/$H${\alpha}$. In both ${\hbox{[S\,{\sc ii}]}}$ \& ${\hbox{[O\,{\sc i}]}}$-BPT diagrams, high-ionization Seyfert galaxies$/$regions and LINERs have been shown to locate on two distinct sequences. Since the S$/$N of ${\hbox{[S\,{\sc ii}]}}$ is much higher than ${\hbox{[O\,{\sc i}]}}$$\lambda6100$ in most spectra, we choose to use the ${\hbox{[S\,{\sc ii}]}}$-BPT diagram to separate LINER and Seyfert regions.\\
\indent Figure \ref{6} shows the ${\hbox{[S\,{\sc ii}]}}$-BPT diagram for the narrow core (top row) and broad wing (bottom row) components. It is clear that the central region within a projected radius of $\sim$4${''}$ (2.4 kpc) is dominated by LINER emission not only for the narrow core but also for most of the broad wing components, while the outer region is dominated by Seyfert-like emission. This result is in agreement with what is shown in Figure \ref{1}(e) and (f). In this work, we define the region within the black polygon as LINER, see Figure \ref{6}(b).

\subsection{Dust attenuation}

\begin{figure*}
	\centering
		\includegraphics[width=0.7\textwidth]{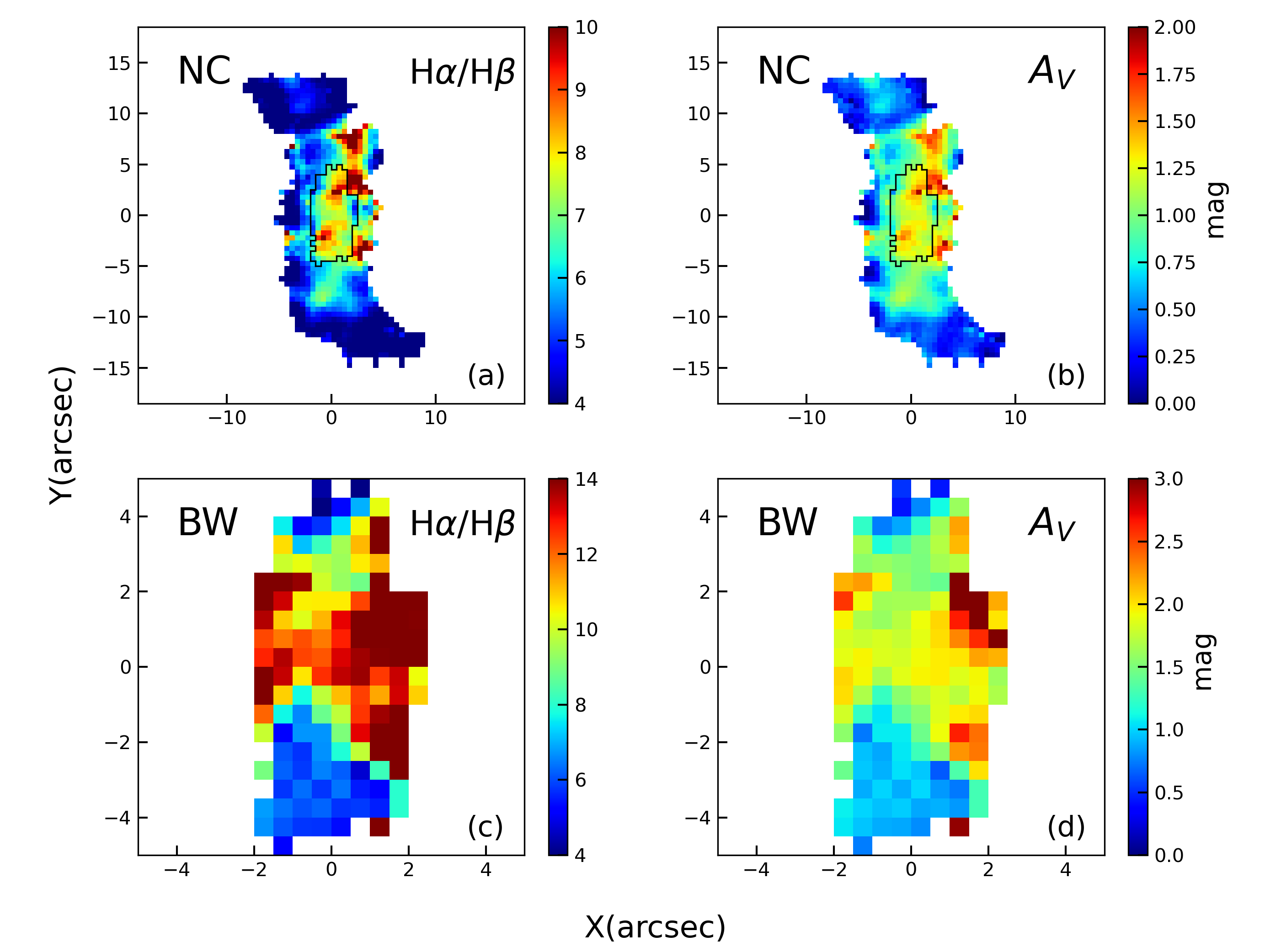}
	\caption{Maps of H${\alpha}$$/$H${\beta}$ and the $V$-band dust attenuation $A_{V}$ derived from the Balmer decrement. (a)$\&$(b) The H${\alpha}$$/$H${\beta}$ $\&$ $A_{V}$ maps for the narrow core component. Similar to Figure \ref{4}, the black polygon marks the region which requires double Gaussian model.
    (c)$\&$(d) The H${\alpha}$$/$H${\beta}$ $\&$ $A_{V}$ maps for the broad wing component. 
    }
	\label{7}
\end{figure*}

Using the flux ratio of H${\alpha}$ $\&$ H${\beta}$ under Case B assumption for temperature $T$ $\sim$ 10,000 K, we correct the dust attenuation according to the following equation: 
\begin{eqnarray} 
F_{\lambda} = F_{\lambda,0} \times 10^{-0.4k(\lambda)E(B-V)},
\label{e1}
\end{eqnarray} 
where $F_{\lambda}$ and $F_{\lambda,0}$ are the observed and intrinsic radiation intensity, $k(\lambda)$ is the Galactic dust extinction curve \citep{2001PASP..113.1449C} with ${R_{V}}$ = 3.1, and color excess $E(B-V)$ = 0.934 $\times$ $\ln$[($\emph{F}_\mathrm{{H\alpha}}$$/$$\emph{F}_\mathrm{{H\beta}}$)].
The $V$-band extinction value in magnitudes is estimated as $A_{V}$ = $E(B-V)$ $\times$ ${R_{V}}$ \citep{1989ApJ...345..245C}. Case B assumes that a nebula is considered optically thick to all Lyman emission lines with wavelength shorter than Ly$\alpha$, in another word these photons are absorbed and re-emitted as a combination of Ly$\alpha$ and higher order emission lines, i.e. Balmer lines \citep{2006agna.book.....O, 2012A&A...542A..30M, 2012MNRAS.419.1402G}.\\ 
\indent Figure \ref{7} shows the maps of H$\alpha$/H$\beta$ and $A_{V}$ for the narrow core (top) and broad wing (bottom) components.
For the narrow core component, the central LINER regions have 0.7 < $A_{V}$ < 1.5 mag with a median value of 1.1 mag, while the Seyfert regions have 0.2 < $A_{V}$ < 1.1 mag with a median value of 0.6 mag. For the broad wing component, it has 0.6 < $A_{V}$ < 2.5 mag with a median value of 1.7 mag. The broad wing component exhibits stronger dust attenuation compared to the narrow core component, since the narrow core component traces the narrow line regions (NLRs), while the broad wing component is originated within NLRs.

\subsection{AGN luminosity} 
\label{4.5}

\begin{table*}
	\centering
		\includegraphics[width=0.65\textwidth]{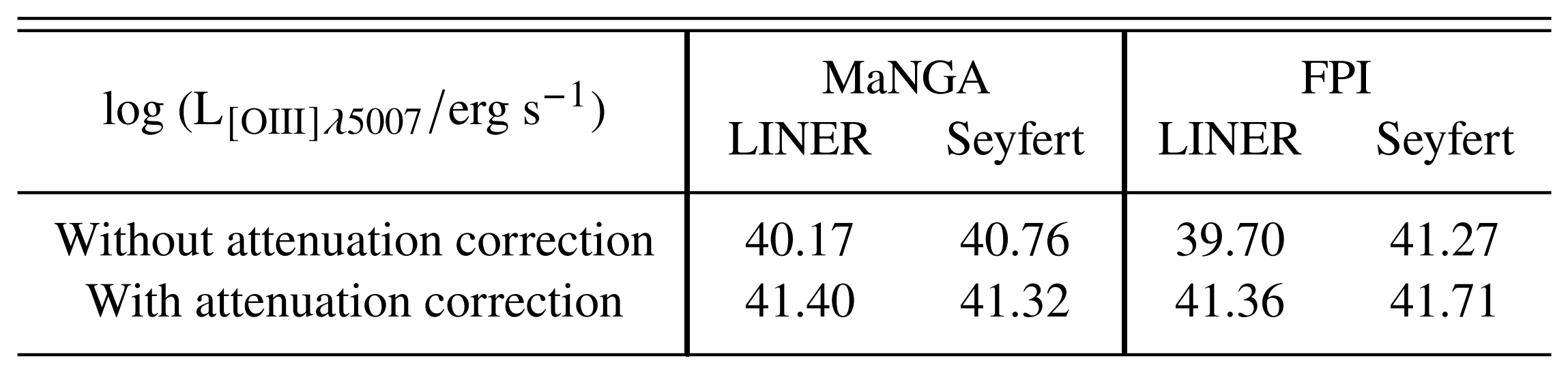}
	\caption{Logarithmic of ${\hbox{[O\,{\sc iii}]}}$$\lambda$5007 luminosity measured from MaNGA and FPI. 
    Column (2): The results are calculated from MaNGA.
    Column (3): The results are calculated from FPI observation. The separation between Seyfert $\&$ LINER regions is shown in Figure \ref{6}(b).
   }
	\label{tab:t1}
\end{table*}

\begin{figure*}
\centering
\includegraphics[width=0.8\textwidth]{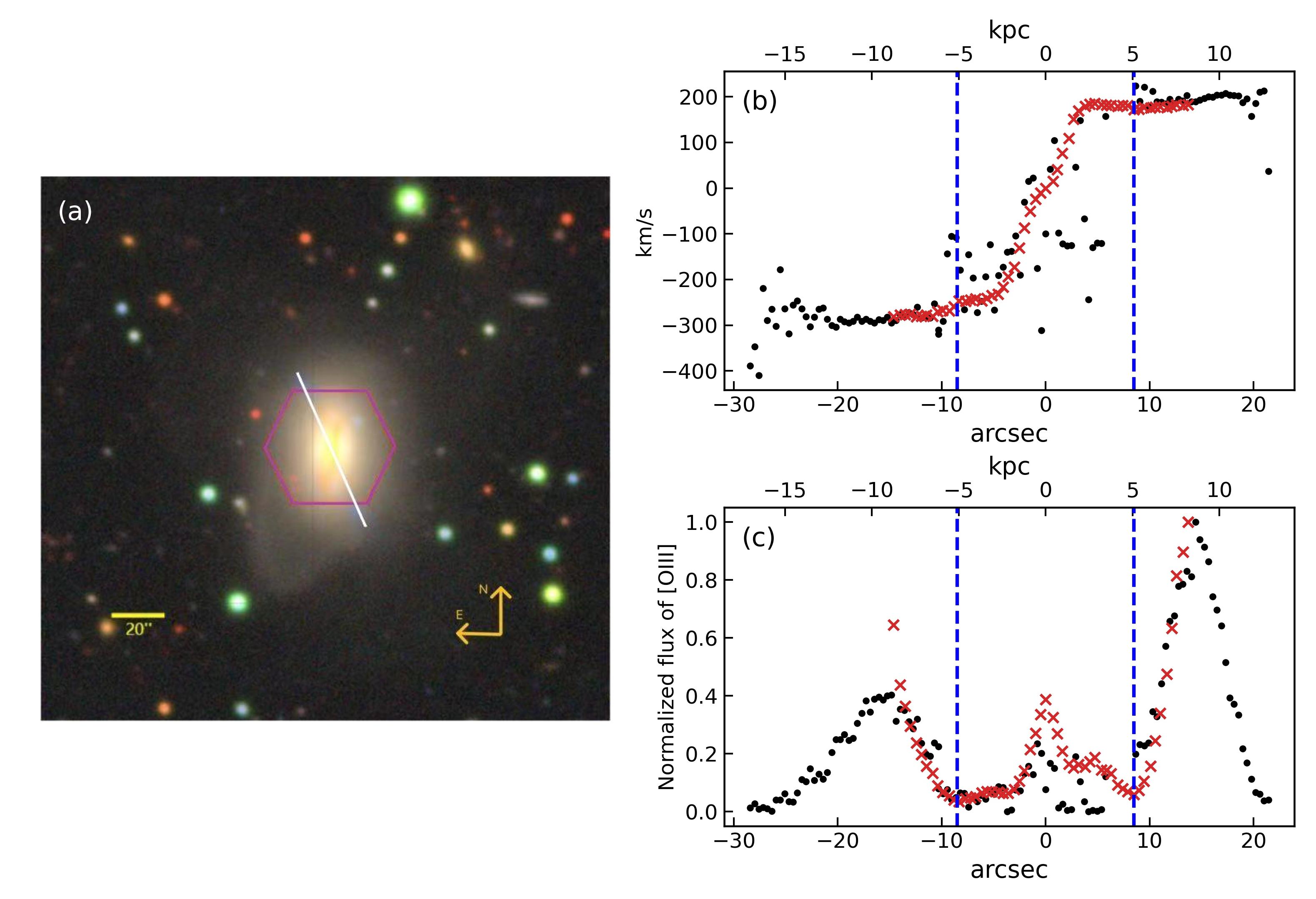}
\caption{\label{9}(a) The DESI image with the white line showing the position of the slit along the two emission line nebulae in blue. (b) The ionized-gas velocities traced by ${\hbox{[O\,{\sc iii}]}}$$\lambda$5007 along the silt.
(c) The normalized ${\hbox{[O\,{\sc iii}]}}$$\lambda$5007 flux along the slit. 
Black dots $\&$ red crosses are measured from APO and MaNGA observation, respectively.
The vertical blue lines mark the inner boundary of the two emission nebulae with enhanced ${\hbox{[O\,{\sc iii}]}}$$\lambda$5007 flux.
}
\label{8}
\end{figure*}

In this section, we calculate the ${\hbox{[O\,{\sc iii}]}}$$\lambda$5007 luminosity for the LINER and Seyfert regions according to the division in Figure \ref{6}(b). We use equation (1) for dust attenuation correction. The observed ${\hbox{[O\,{\sc iii}]}}$$\lambda$5007 luminosity ($L$$_{\mathrm{[OIII]}}^{\mathrm{obs}}$) from MaNGA data is 10$^{40.17}$ erg s$^{-1}$ and 10$^{40.76}$ erg s$^{-1}$ for LINER and Seyfert regions, respectively. The ${\hbox{[O\,{\sc iii}]}}$$\lambda$5007 luminosity after attenuation correction ($L$$_{\mathrm{[OIII]}}^{\mathrm{inc}}$) is 10$^{41.40}$ erg s$^{-1}$ and 10$^{41.32}$ erg s$^{-1}$ for LINER and Seyfert regions, respectively. We note that the MaNGA bundle only covers part of the two emission line nebulae with enhanced ${\hbox{[O\,{\sc iii}]}}$$\lambda$5007 EQW in the northeast and southwest, which are shown as blue blobs in Figure \ref{8}(a). Thus, the ${\hbox{[O\,{\sc iii}]}}$$\lambda$5007 luminosity 
for the Seyfert regions is underestimated.\\
\indent We observe this galaxy with 3.5m telescope at the Apache Point Observatory (APO) and obtain the long-slit spectra with an observable wavelength coverage of 3874$\sim$7823{\AA}. Figure \ref{8}(a) shows the DESI image with the white line marking the slit position. Figure \ref{8}(b)$\&$(c) show the ionized-gas velocities traced by ${\hbox{[O\,{\sc iii}]}}$$\lambda$5007 emission line and normalized ${\hbox{[O\,{\sc iii}]}}$$\lambda$5007 flux along the slit, respectively. Black dots $\&$ red crosses are measured from APO and MaNGA observations, respectively. The vertical blue lines mark the inner boundary of the two emission nebulae with enhanced ${\hbox{[O\,{\sc iii}]}}$$\lambda$5007 flux. It is clear that APO and MaNGA observations give consistent results in both line-of-sight velocity and ${\hbox{[O\,{\sc iii}]}}$$\lambda$5007 flux. The emission line ratios measured from APO data show that the two emission line nebulae outside the MaNGA bundle are also located at the Seyfert region, the median values of log(${\hbox{[O\,{\sc iii}]}}$$\lambda5007$$/$H${\beta}$) is 0.95 and log(${\hbox{[S\,{\sc ii}]}}$$\lambda\lambda6717, 6731$$/$H${\alpha}$) is $-$0.35, shown as the blue star in Figure \ref{6}(a).\\  
\begin{figure*}
	\centering
		\includegraphics[width=\textwidth]{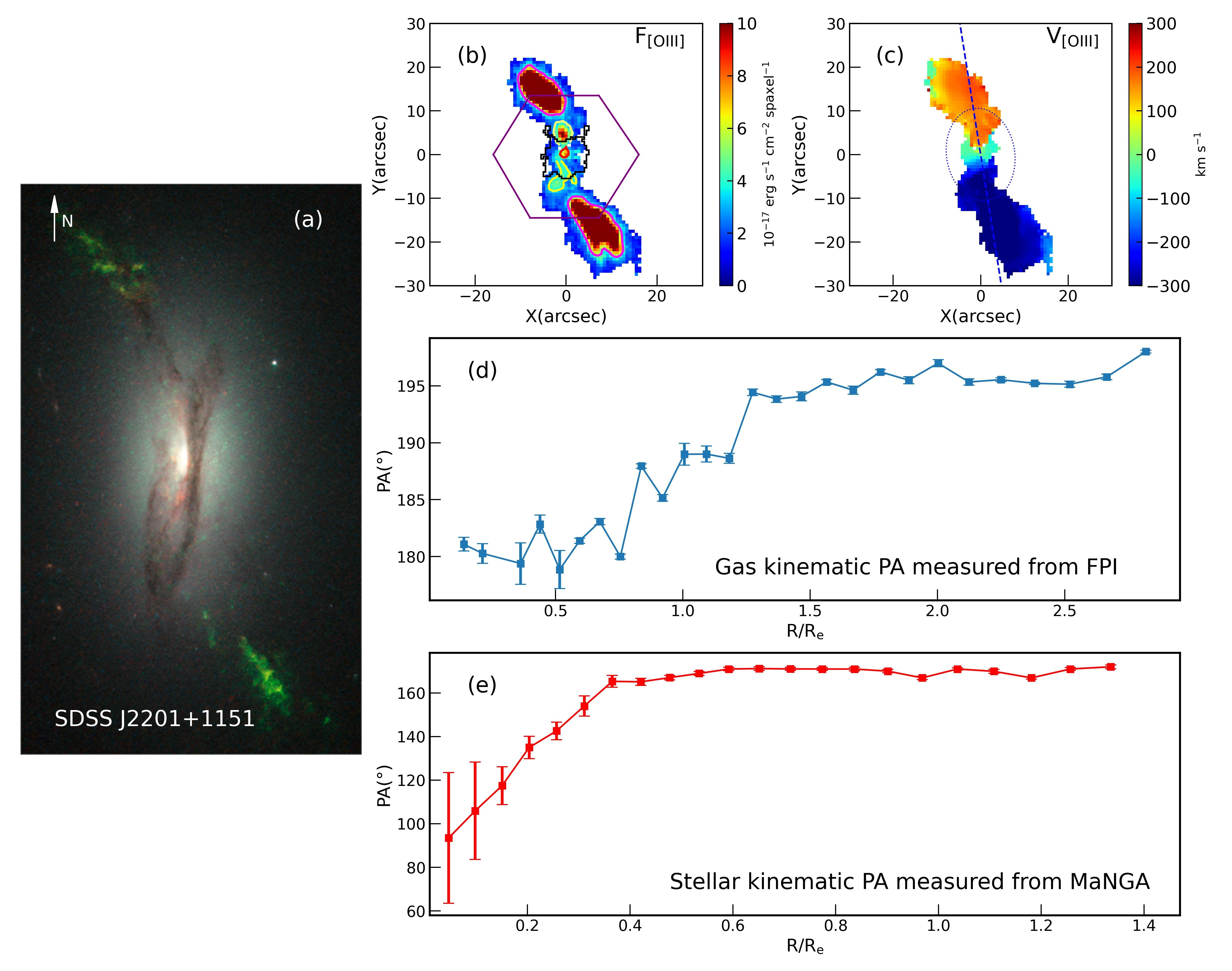}
	\caption{(a) HST image of SDSS J2201+1151 which compares the continuum and emission line structure. Red $\&$ green are the continuum and ${\hbox{[O\,{\sc iii}]}}$$\lambda$5007 emission, respectively. 
    (b) The FPI ${\hbox{[O\,{\sc iii}]}}$$\lambda$5007 flux map. The purple hexagon shows the MaNGA bundle, while pink, yellow, and red contours mark the regions with enhanced ${\hbox{[O\,{\sc iii}]}}$$\lambda$5007 flux. The LINER regions from MaNGA observation are marked by the black polygon. 
    (c) The gas velocity field traced by ${\hbox{[O\,{\sc iii}]}}$$\lambda$5007 emission line measured from the FPI observations.
    The blue dotted ellipse shows 1$R_{e}$ region of this galaxy and the blue dashed line marks the kinematic major axis of the ellipse.
    (d)$\&$(e) Kinematic position angle as a function of radius measured from the gas and stellar velocity field where the error bars show $\pm1$$\sigma$ scattering region, respectively. 
    }
	\label{9}
\end{figure*}
\indent To obtain robust ${\hbox{[O\,{\sc iii}]}}$$\lambda$5007 luminosity for the Seyfert regions, we use the ${\hbox{[O\,{\sc iii}]}}$$\lambda$5007 observation from the 6m BTA in its scanning FPI mode firstly presented in \citet{2015AJ....149..155K}. Figure \ref{9}(a) shows the HST image which compares the continuum and emission line structure of this galaxy. Red $\&$ green are the continuum and ${\hbox{[O\,{\sc iii}]}}$ emission, respectively. Figure \ref{9}(b)$\&$(c) are the ${\hbox{[O\,{\sc iii}]}}$$\lambda$5007 flux and line-of-sight velocity maps observerd by FPI. The purple hexagon in Figure \ref{9}(b) marks the position of the MaNGA bundle, while pink, yellow, and red contours mark regions with enhanced ${\hbox{[O\,{\sc iii}]}}$$\lambda$5007 flux at different radii. The LINER regions observed from MaNGA are marked by the black polygon. In Figure \ref{9}(c), the blue dotted ellipse shows 1$R_{e}$ region of this galaxy and the blue dashed line marks the kinematic major axis of the ellipse.\\ 
\indent The observed ${\hbox{[O\,{\sc iii}]}}$$\lambda$5007 luminosity measured from FPI is 10$^{39.70}$ erg s$^{-1}$ and 10$^{41.27}$ erg s$^{-1}$ for LINER $\&$ Seyfert regions, respectively. However, we lack the measurements of dust attenuation for the two emission line nebulae outside the MaNGA bundle. We assume that the two emission line nebula outside the MaNGA bundle have similar dust attenuation values as that measured from the long slit of APO in the Seyfert regions outside the MaNGA bundle. This gives a intrinsic luminosity of $L$$_{\mathrm{[OIII]}}^{\mathrm{inc}}$ =  10$^{41.71}$ erg s$^{-1}$ for Seyfert region. In Table \ref{tab:t1}, we list the ${\hbox{[O\,{\sc iii}]}}$$\lambda$5007 luminosity measured from MaNGA and FPI in detail.\\ 
\indent The ${\hbox{[O\,{\sc iii}]}}$$\lambda$5007 luminosity of the Seyfert regions is a factor of 37 (2) higher than the LINER regions without (with) dust attenuation correction, indicating the current AGN activity is insufficient to power the two emission line nebulae.
One natural explanation of this discrepancy is that the AGN activity has faded with time. Given a projected distance of $\sim$2.4 kpc between the nucleus and the boundary of the LINER regions defined in Figure \ref{6}(b), we suggest that the AGN activity decreases at least $\sim$8 $\times$ 10$^3$ yrs ($\sim$2.4 kpc/light-speed) ago.

\subsection{Kinematic position angle of the gas and stellar components}
In this section, we fit the kinematic position angle (PA) of the   
${\hbox{[O\,{\sc iii}]}}$$\lambda$5007 velocity field in Figure \ref{9}(c) and the stellar velocity field in Figure \ref{5}(a) as a function of radius based on the established method \citep{2006MNRAS.366..787K}. The kinematic PA is defined as the counter-clockwise angle between the north and a line that bisects the velocity field of gas, measured on the approaching side. Figure \ref{9}(d) shows the kinematic PA as well as $\pm$1$\sigma$ error range of the gas component as a function of radius. The PA increases from about 180° at the galaxy center to about 200° at 1.5$R_e$. Figure \ref{9}(e) shows the kinematic PA as well as $\pm$1$\sigma$ error range of the stellar component as a function of radius. It is clear that there is an obvious variation of stellar PA within $\sim$0.4$R_{e}$. The ellipse with 0.4$R_{e}$ kinematic major axis is marked in Figure \ref{5}(a). Comparing the gradients of gas $\&$ stellar PAs, we find although the warping direction of gas and stellar disks is consistent, the median of the position angle difference is as large as $\sim$50° within 0.4$R_{e}$, and decreases to $\sim$25° over this radii.

\section{DISCUSSION}
\subsection{Comparison of SDSS J2201+1151 properties with previous studies} \label{dis1} 
In this section, we compare our results from Section \ref{3} with previous studies, including kinematics of the broad wing component, AGN luminosity, and kinematics position angle of the gas component.\\
\indent Based on a large sample of $\sim$39,000 type 2 AGNs at $z$ < 0.3 from the MPA-JHU Catalog of the SDSS Data Release 7 \citep{2009ApJS..182..543A}, \citet{2016ApJ...817..108W} perform a detailed analysis on ${\hbox{[O\,{\sc iii}]}}$$\lambda$5007 emission line structure, finding that in $\sim$43.6$\%$ AGNs, the ${\hbox{[O\,{\sc iii}]}}$$\lambda$5007 emission line cannot be model by a single Gaussian model. Two Gaussian components (a narrow core plus a broad wing) are required to model ${\hbox{[O\,{\sc iii}]}}$$\lambda$5007 emission line, suggesting that the broad wing component originated from ionized gas outflows are prevalent among type 2 AGNs. \citet{2017ApJ...845..131K} suggest that strong outflows are ubiquitous in galaxies with $\sigma$$_{\mathrm{H\alpha}}$$/$$\sigma$$_{\star}$ > 1.4, where $\sigma$$_{\mathrm{H\alpha}}$ $\&$ $\sigma$$_{\star}$ is the velocity dispersion of the H$\alpha$ emission and stellar component, respectively. We closely follow the method of \citet{2017ApJ...845..131K} to estimate the velocity dispersion from ${\hbox{[O\,{\sc iii}]}}$$\lambda$5007 profile, finding a median dispersion of $\sim$300 km s$^{-1}$ within the central 3" region (the aperture of a single fiber used in the early stage of SDSS survey), which is 1.5 times larger than the median of the stellar velocity dispersion $\sim$200 km s$^{-1}$ in the same region, suggesting the existence of strong outflow in galaxy SDSS J2201+1151. Moreover, the velocity dispersion of the broad wing component ranging from  400$\sim$500 km s$^{-1}$ is suggested to originate from shock generated by the outflow interaction with the interstellar medium \citep{2019MNRAS.487.4153D}.\\
\indent \citet{2012MNRAS.420..878K} calculate the FIR AGN luminosity and the luminosity required for the EELRs ionization in SDSS J2201+1151, finding that the luminosity required for the EELRs ionization is 3.4 times larger than the FIR AGN luminosity, suggesting the extended cloud can not be ionized by
an obscured AGN in SDSS J2201+1151. The most direct interpretation is that the EELRs are ionized by AGN that has faded over the light-travel time between the ionized gas and nucleus, which is consistent with our result that the bolometric luminosity in the LINER region is 2 times lower than the Seyfert regions.\\
\indent The gas kinematic PA measured from FPI increases from about 180° at the galaxy center to about 200° at 1.5$R_e$ shown in Figure \ref{9}(d). 
The $\sim$20° change of PA from the inner to the outer region is consistent with \citet{2015AJ....149..155K} in that they model this galaxy with a differential processing disk, finding the existence of a warped disk tilted by $\sim$23 degrees to the stellar disk of the host galaxy.

\subsection{SDSS J2201+1151: A fading AGN}
SDSS J2201+1151 shows LINER-like ionization within its central $\sim$2.4 kpc while the outskirts show Seyfert-like emission. 
The ${\hbox{[O\,{\sc iii}]}}$$\lambda$5007 luminosity of the Seyfert region is about 2 times higher than the LINER region after dust attenuation correction, suggesting that this galaxy as a fading AGN. In this section, we further confirm this galaxy as a fading AGN by comparing its multi-band luminosities.\\
\indent It is well-known that differences in the energy budget between different AGN components, like the accretion disk, dust torus, and optical ionizing regions, can be interpreted as a hint on AGN evolution \citep{2015MNRAS.451.2517S, 2019ApJ...870...65I, 2019ApJ...883L..13I, 2020ApJ...905...29E}. As we know, the X-ray emission traces the nuclear source luminosity originates from $\sim$10$^{-2}$ pc-scale or timescale of 3 $\times$ $10^{-2}$ yrs \citep{2007A&A...462..581H}. Mid-infrared (MIR) radiation is dominated by the obscuring dust located few parsecs away from the nucleus \citep{2017NatAs...1..679R}. ${\hbox{[O\,{\sc iii}]}}$$\lambda$5007 luminosity traces the narrow emission line region which is located at kpc scale \citep{2017ApJ...835..256K}. The closer the structure to the galaxy center, the shorter the timescale of the evolution. In a fading scenario, we would expect the luminosity from the NLR to be higher than that from obscuring dust, while the accretion disk current bolometric luminosity being the lowest among them. \\
\indent \citet{2020ApJ...905...29E} developed a method to select fading/rising AGNs by applying the linear relations between radiations of several AGN structures at different observing bands. Figure \ref{10}(a) shows the best fit linear relation (black line) between 2-10keV X-ray luminosity and the attenuation corrected ${\hbox{[O\,{\sc iii}]}}$$\lambda$5007 luminosity in logarithmic space (see their Figure 1). Figure \ref{10}(b) shows the 2-10keV X-ray luminosity versus the MIR 12$\mu$m continuum luminosity, again the black line is the best-fit linear relation. 
In both panels, the gray shadowed region indicates the  $\pm$1$\sigma$ scattering region. AGNs located up the $+$1$\sigma$ boundary are defined as rising AGNs while fading AGNs are located down the $-$1$\sigma$ boundary. We estimate 2-10keV X-ray luminosity from XMM-Newton and MIR 12$\mu$m continuum luminosity from WISE. SDSS J2201+1151 is shown as red stars in both panels of Figure \ref{10} and it is clear that this galaxy deviates far from the linear relation, clearly located at the fading AGN region, which is consistent with previous {${\hbox{[O\,{\sc iii}]}}$$\lambda$5007 luminosity measurements of Seyfert and LINER regions in Section \ref{4.5}.\\

\begin{figure*}
\centering
\includegraphics[width=0.8\textwidth]{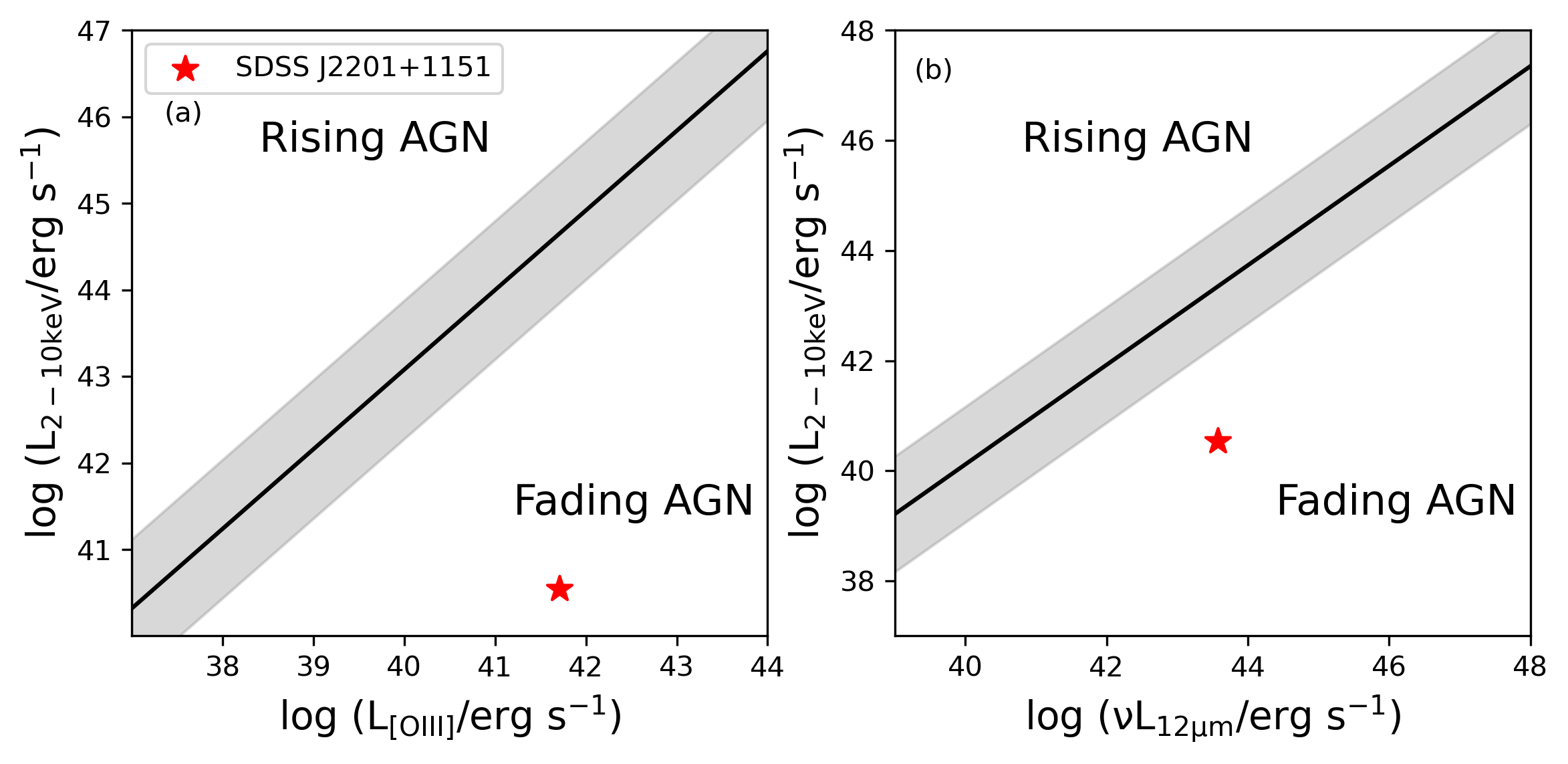}
\caption{\label{10} (a) 2-10keV X-ray luminosity versus the dust attenuation corrected ${\hbox{[O\,{\sc iii}]}}$$\lambda$5007 luminosity (both in logarithmic scale). (b) 2-10keV X-ray luminosity versus the MIR 12$\mu$m continuum luminosity in logarithmic scales. The black lines show the best-fit linear with the gray shadowed region indicating the $\pm$1$\sigma$ scattering region \citep{2020ApJ...905...29E}. SDSS J2201+1151 is marked as red stars in both panels.}
\end{figure*}

\subsection{Episodic AGN outbursts}
Figure \ref{9}(b) shows several regions with enhanced ${\hbox{[O\,{\sc iii}]}}$$\lambda$5007 flux marked by contours with different colors, there are obvious flux gaps between these regions. The central region marked by the red contour is characterized as LINER and the outer regions marked by the pink contours are dominated by Seyfert. However, the regions in-between marked by the yellow contours on the north side is dominated by Seyfert while half of the south yellow contour is located within the LINER region. This discontinuity is also found in ${\hbox{[O\,{\sc iii}]}}$$\lambda$5007 EQW map, see the black contours in Figure \ref{2}(a).

One possible explanation for the discontinuity in the distribution of ${\hbox{[O\,{\sc iii}]}}$$\lambda$5007 flux and EQW is episodic AGN outbursts. The tidal feature in Figure \ref{1}(b) indicates SDSS J2201+1151 is a merger remnant. The merger process kick-started the central engine to the quasar phase which ionized gas composed of tidal debris, resulting in the regions marked by pink contours. The largest projected distance from the pink contours to the galactic center is about 15 kpc, corresponding to an AGN burst that happened at least $\sim$5 $\times$ 10$^4$ yrs ($\sim$15 kpc/light-speed) ago.

The regions marked by yellow contours correspond to a second AGN outburst due to the merge remnant gas inflow which occurred at least $\sim$1.3 $\times$ 10$^4$ yrs ago corresponding to a projected distance of $\sim$4 kpc between the nucleus and the farthest end of the yellow contours.

\section{Conclusions}
We identify a fading AGN SDSS J2201+1151 from MaNGA MPL11. We use the double Gaussian model (a narrow core and a broad wing) to describe the profile of the emission line spectra for the spaxel within the black polygon marked in Figure \ref{6}(b), while only a single Gaussian model (a narrow core) is applied to fit each emission line for the outer region and analyze the detailed properties of this galaxy:\\
\indent i) The central region within a projected radius of $\sim$2.4 kpc is dominated by LINER emission not only for the narrow core but also for the broad wing components, while the outer region is dominated by Seyfert-like line ratios. We obtain robust ${\hbox{[O\,{\sc iii}]}}$$\lambda$5007 luminosity for the Seyfert and LINER regions, finding ${\hbox{[O\,{\sc iii}]}}$$\lambda$5007 luminosity of the Seyfert region is 37 (2) times higher than that of the LINER region without (with) dust attenuation correction, indicating that the AGN activity decreases at least $\sim$8 × 10$^3$ yrs ago.\\
\indent ii) The narrow core component appears to be co-rotating with the stellar component, while the kinematics of the
broad wing component seems different from the stellar component and much more complicated. The median velocity dispersion is $\sim$100 km s$^{-1}$ for the narrow core component and $\sim$300 km s$^{-1}$ for the broad wing component. 
The velocity dispersion of the broad wing component within the central $\pm$2${''}$ region is up to 400$\sim$500 km s$^{-1}$, which is related to a wind outflow. The broad wing component exhibits stronger dust attenuation compared to the narrow core component, since the narrow core component traces NLRs, while the broad wing component is originated within NLRs.\\
\indent iii) There are obvious gaps between several regions with enhanced ${\hbox{[O\,{\sc iii}]}}$$\lambda$5007 flux and EQW. One possible explanation for the discontinuity in the distribution of ${\hbox{[O\,{\sc iii}]}}$$\lambda$5007 flux and EQW maps is episodic AGN outbursts.\\
\indent iv)We use well-known linear relations between luminosities in logarithmic space of several AGN components at different wavelengths. These relations are based on the assumption that when the AGNs have no variations in luminosity, the different AGN components trace the same bolometric luminosity. Outliers of the relations imply variability in AGN activity. SDSS J2201+1151 deviates far from the linear relation, obviously located at the fading AGN region.\\
\indent v) The kinematic position angle of ionized gas increases from about 180° at the galaxy center to about 200° at 1.5$R_e$ and there is an obvious variation of stellar PA within $\sim$0.4$R_{e}$. Comparing the gradients of gas $\&$ stellar PAs, we find although the warping direction of gas and stellar disks is consistent, the median of the position angle difference is as large as $\sim$50° within 0.4$R_{e}$, and decreases to $\sim$25° over this radii. The large misalignment between gas $\&$ stellar components within 0.4$R_{e}$ suggests the influence of external process like merger on this galaxy.\\
\indent The tidal features in DESI image and star-gas misalignment list above suggest this galaxy is a merger remnant. Combining all these observational results, we confirm this is a fading AGN and provide a scenario that the merger process kick-started the central engine to quasar phase which ionized gas composed of tidal debris. These episodic AGN outbursts could be the result of the galaxy merger that drives discrete accretion events onto the supermassive black hole. 

\section*{Acknowledgements}

M.Y.C. acknowledges support from the National Natural Science Foundation of China (NSFC grants 12333002, 11733002), the China Manned Space Project with NO. CMS-CSST-2021-A05 as well as the China Manned Space Project (the second-stage CSST science project: “Investigation of small-scale structures in galaxies and forecasting of observations”. M. B. acknowledges support from the National Natural Science Foundation of China (NSFC grants 12303009). A.M. and D.B. acknowledges partial support from the Russian Science Foundation (grant No. 22-12-00080). This research is partly based on observations obtained with the Apache Point Observatory 3.5-meter telescope, which is owned and operated by the Astrophysical Research Consortium.

Funding for the Sloan Digital Sky Survey IV has been provided by the Alfred P. Sloan Foundation, the U.S. Department of Energy Office of Science, and the Participating Institutions. SDSS-IV acknowledges support and resources from the Center for High-Performance Computing at the University of Utah. The SDSS web site is www.sdss.org.

SDSS-IV is managed by the Astrophysical Research Consortium for the Participating Institutions of the SDSS Collaboration including the Brazilian Participation Group, the Carnegie Institution for Science, Carnegie Mellon University, the Chilean Participation Group, the French Participation Group, Harvard-Smithsonian Center for Astrophysics, Instituto de Astrofísica de Canarias, The Johns Hopkins University, Kavli Institute for the Physics and Mathematics of the Universe (IPMU) / University of Tokyo, Lawrence Berkeley National Laboratory, Leibniz Institut für Astrophysik Potsdam (AIP), MaxPlanck-Institut für Astronomie (MPIA Heidelberg), Max-PlanckInstitut für Astrophysik (MPA Garching), Max-Planck-Institut für Extraterrestrische Physik (MPE), National Astronomical Observatories of China, New Mexico State University, New York University, University of Notre Dame, Observatário Nacional / MCTI, The Ohio State University, Pennsylvania State University, Shanghai Astronomical Observatory, United Kingdom Participation Group, Universidad Nacional Autónoma de México, University of Arizona, University of Colorado Boulder, University of Oxford, University of Portsmouth, University of Utah, University of Virginia, University of Washington, University of Wisconsin, Vanderbilt University, and Yale University.

\section*{Data Availability}

The data underlying this article will be shared on reasonable request
to the corresponding author.



\bibliographystyle{mnras}
\bibliography{referee} 

\bsp	
\label{lastpage}
\end{document}